\documentclass[aps,pra,twocolumn,groupedaddress]{revtex4}
\usepackage{amsmath}
\usepackage{amssymb}
\usepackage{graphicx}
\usepackage{bm}
\usepackage{appendix}

\newcommand{\non}{\nonumber}
\newcommand{\sbp}{\subparagraph*{}}
\newcommand{\be}{\begin{equation} }
\newcommand{\ee}{\end{equation}}
\newcommand{\bea}{\begin{eqnarray} }
\newcommand{\eea}{\end{eqnarray} }
\newcommand{\sg}{\sigma}
\newcommand{\gm}{\gamma}
\newcommand{\hb}{\hbar}
\newcommand{\alp}{\alpha}
\newcommand{\og}{\omega}
\newcommand{\Og}{\Omega}
\newcommand{\lb}{\lambda}

\newcommand{\tbf}{\textbf}
\newcommand{\bds}{\boldsymbol}

\newcommand{\dt}{\delta}
\newcommand{\ep}{\epsilon}

\newcommand{\comment}[1]{}

\newcommand{\vep}{\varepsilon}
\newcommand{\half}{\frac{1}{2}}
\newcommand{\ptl}{\partial}

\newcommand{\Tr}{{\rm Tr}}
\newcommand{\Rl}{{\rm Re}}
\newcommand{\diag}{{\rm diag}}

\newcommand{\im}{{\rm i}}
\renewcommand\Re{\mathfrak{Re}\,}

\begin{document}

\title{Entanglement of a two-particle Gaussian state interacting with a heat bath}

\author{Anne Ghesqui\`{e}re}
\email[]{aghesquiere@stp.dias.ie}
%\thanks{}
\altaffiliation{Department of Mathematical Physics, NUI Maynooth, Maynooth, Co. Kildare, Ireland}
\affiliation{School of Theoretical Physics, Dublin Institute for Advanced Studies, 10 Burlington Road, Dublin 4, Ireland}

\author{Tony C. Dorlas}
\email[]{dorlas@stp.dias.ie}
%\thanks{}
\affiliation{School of Theoretical Physics, Dublin Institute for Advanced Studies, 10 Burlington Road, Dublin 4, Ireland}

\date{\today}

\pacs{}
%\keywords{}


\begin{abstract}
The effect of a thermal reservoir is investigated on a bipartite Gaussian state. We derive a pre-Lindblad master equation in the non-rotating wave approximation for the system. We then solve the master equation for a bipartite harmonic oscillator Hamiltonian with entangled initial state. We show that for strong damping the loss of entanglement is the same as for freely evolving particles. However, if the damping is small, the entanglement is shown to oscillate and eventually tend to a constant nonzero value.
\end{abstract}

\maketitle


\section*{Motivation}

\sbp Entanglement is one of quantum mechanics' most fascinating features. It was first described in a celebrated paper by Einstein, Podolsky and Rosen \cite{EPR:1935} but owes its name to Schr\"odinger \cite{Schrod:1935}, who investigated its broader significance for the measurement question. It has taken on enhanced significance in quantum information. In this regard, the fragility of entanglement when the system is subjected to ``outside'' influence is of even greater importance. In the current work, we study a bipartite system with a Gaussian wave function. The state is prepared such that it is entangled, then shared between two parties who let their respective particle evolve either freely or interacting via a harmonic potential, but interacting with its own environment or heat bath. We study the resulting loss of entanglement between the particles. To do so, we first give a simple derivation of a pre-Lindblad non-rotating-wave master equation, \cite{Gard:NRWME, Gard:2000}, starting with the Quantum Langevin Equation as derived in \cite{FLO'C:QLE1988} and using a simple perturbation method as in \cite{Sakuya:2009}.

The loss of entanglement in a system interacting with an environment is a well-known phenomenon. It has been studied in various systems, see e.g. \cite{YE:2003, YE:2004, YE:2006, PE:2001, Diosi:2003, RM:2006}, where it was found that there is often a sharp loss of entanglement when compared to a decoherence time scale, which has been termed entanglement sudden-death (E.S.D.). These studies are mainly in the context of qubits and the Rotating Wave Approximation (R.W.A.) however, whereas this work presents a study of E.S.D. in a continuous-variables setting and uses the Non-Rotating-Wave (N.R.W) approximation. Note that the master equation obtained in the N.R.W approximation is not of the Lindblad form \cite{Lindblad:1976}, hence could in principle have unphysical results. However, the unphysical behavior is obtained for low T, and one can easily check for the validity of the density matrix by checking its positive semi-definiteness. On the other hand, the N.R.W master equation often works better for systems which are strongly coupled to the environment \cite{Gard:NRWME}.

In \cite{Ficek:2006}, Ficek and Tan\'as study a system of two qubits coupled to a radiation field where they allow spontaneous decay of the atoms. They show that the entanglement vanishes but then is revived twice. In \cite{Ficek:2008}, the authors study the emergence of entanglement between two initially non-entangled
qubits due to spontaneous emission, provided both atoms are initially excited and in the asymmetric state. Their results suggest that an interaction between two particles which are initially entangled can delay the vanishing of the entanglement and even revive it, or create entanglement between two initially non-entangled particles. We introduce a harmonic potential with frequency $\og_0$ as the interaction between the particles in our system and examine the dynamics of the entanglement. We show that entanglement revival can occur depending on the strength of the damping, i.e. how strong the coupling $\gm$ is with respect to the
oscillator's frequency. We show that if the damping is small ($\gm < 2 \sqrt{2} \og_0 $), the entanglement oscillates towards a central value and does not vanish asymptotically.

In Section~\ref{framework} we recall the Langevin equation and the main steps in the derivation of the master equation. We then establish in Section~\ref{GaussState} the formalism used to describe Gaussian states and the particular measure for entanglement we use. Section~\ref{result} illustrates E.S.D. while Section~\ref{bipharmpot} contains the main results of this paper. Section~\ref{conclusion} contains some concluding remarks.

\section{Framework\label{framework}}

The derivation of the master equation is detailed in more detail in Appendix~\ref{derivmasteq} and will only be succintly presented in the following. The derivation is easiest for one-particle but generalises just as easily to the case of two-particles, each coupled to its own environment. We consider a heat bath modelled by independent oscillators coupled harmonically to the particle \cite{FLO'C:QLE1988}. The corresponding Hamiltonian has the form 
\begin{equation}
H = \frac{p^2}{2m} + V(x) + \frac{1}{2} \sum_j \left\{
\frac{p_j^2}{m_j} + m_j \og_j ^2 (q_j - x)^2 \right\}
\end{equation}
Solving the Heisenberg equations of motion for $q_j$ yields the  Quantum Langevin Equation
\begin{eqnarray}
m {\ddot x} + \int_{-\infty} ^t \mu (t-t') {\dot x}(t')\, dt' + V'(x) &=& \xi(t)
\end{eqnarray}
where the dot denotes the derivative with respect to time and the prime that with respect to $x$. $\mu(t)$ and $\xi(t)$ describe the influence of the bath on the system and are known as the \textit{memory function} and the operator-valued \textit{random force} respectively and are expressed explicitly in Appendix~\ref{derivmasteq}.
In the case of a Ohmic heat bath, $\mu(t)$ effectively reduces to $\gm$ and we can write, for a general observable $Y$ of the small system (particle), the Quantum Langevin Equation reads
\begin{equation} \label{qle1} {\dot Y} = \frac{\im}{\hb}
\left[ H_s, Y \right] - \frac{\im}{2\hb} \left[ \left[ x ,Y
\right], \xi (t) \right]_+ + \frac{\im \gm}{2\hb} \left[ \left[
x, Y \right], {\dot x} (t) \right]_+. \end{equation} 
For a system of two particles, each connected to their individual heat bath, we have analogously
\begin{eqnarray} \label{qle}
{\dot Y} &=& \frac{\im}{\hb} \left[ H_s, Y \right] \non 
\\ && - \frac{\im}{2\hb} \left[ \left[ x_1 ,Y \right], \xi_1 (t)
\right]_+ - \frac{\im}{2\hb} \left[ \left[ x_2 ,Y \right],
\xi_2 (t) \right]_+ \non \\ && + \frac{\im}{2\hb} \left[ \left[
x_1 , Y \right] , \gm_1 {\dot x_1} (t) \right]_+ +
\frac{\im}{2\hb} \left[ \left[ x_2 , Y \right] ,\gm_2 {\dot
x_2} (t) \right]_+ \non \\
\end{eqnarray}
The Langevin equation is an equation for the system operators (Heisenberg representation), whereas a master equation is an approximate equation acting on the density operator of the quantum system under study (Schr\"odinger picture). The adjoint equation provides a link between the two formalisms. We define
\begin{equation} \label{trs}
 \Tr_s \left\{ Y(t) \rho \right\} = \Tr_s \left\{ Y \rho(t) \right\} \non
\end{equation}
and obtain the adjoint equation
\begin{eqnarray} \label{adj}
{\dot \rho}_s(t) &=& - \frac{\im}{\hb} \left[ H_s , \rho_s(t) \right] - \frac{\im}{2\hb} \left[ \left[ \xi_1 (t) , \rho_s(t) \right]_+ ,
x_1 \right] \non 
\\ && - \frac{\im}{2\hb} \left[ \left[ \xi_2 (t) , \rho_s(t) \right]_+ , x_2 \right] 
\\ && + \frac{\im \gm_1}{2\hb} \left[ \left[ {\dot x}_1 , \rho_s(t) \right]_+ , x_1 \right] + \frac{\im \gm_2}{2\hb} \left[ \left[ {\dot x}_2 , \rho_s(t)
\right]_+ , x_2 \right] \non 
\end{eqnarray}
In order to derive the master equation let us assume that the bath is large so that we may assume that it stays at thermal equilibrium and that at $t \rightarrow - \infty$, the system and the bath are decoupled so that $\rho_0 (t) = \rho_s (t) \rho_B$. This assumption is critical to the derivation of any master equation. Finally, assuming that the noise is small allows us to write $\xi(t) \rightarrow \ep \xi(t)$. This assumption is not essential to the derivation but allows for a simpler derivation. Applying a perturbation method and tracing over the bath yields the Non-Rotating-Wave master equation for $\rho_s(t)$ 
\begin{eqnarray} \label{mastereqn}
{\dot \rho}(t) &=& -\frac{\im}{\hb} \left[H_s , \rho (t)\right] \non
\\ && + \frac{\im \gm_1}{2\hb} \left[ \left[ {\dot x}_1 , \rho (t) \right]_+ , x_1 \right] 
+ \frac{\im \gm_2}{2\hb} \left[ \left[ {\dot x}_2 , \rho (t) \right]_+ , x_2 \right] \non 
\\ && - \frac{k T_1 \gm_1}{\hb ^2} \left[ \left[ \rho(t) , x_1 \right] , x_1 \right]
 - \frac{k T_2 \gm_2}{\hb ^2} \left[ \left[ \rho(t) , x_2 \right] , x_2 \right]. \non \\
\end{eqnarray}

\section{Gaussian states and the logarithmic negativity \label{GaussState}}

Since the states we will study are Gaussian, we now briefly recall the formalism for Gaussian states \cite{Anders:2003, Eisert:2003, Eisert:2004}. 
\sbp Gaussian states can be completely specified in terms of their first and second moments, described respectively by the displacement vector  
\be d_j=\langle R_j\rangle_{\rho}=\Tr[R_j\rho] \non \ee 
and the covariance matrix
\be \gm_{j,k}=2\,\Re \Tr[\rho( R_j-\langle R_j \rangle_{\rho})(R_k-\langle R_k \rangle_{\rho})] \non \ee 
where $R$ is the vector $R^T= (q_1,p_1; ....; q_n,p_n)$ ; $q_j$ and $p_j$ are the canonical variables of a system of $n$ oscillators with the usual canonical relations written as $\left[ R_j, R_k \right] = \im \hb \sg_{jk}$ and $\sg$ a real skew-symmetric $2n \times 2n$ block matrix given by 
\be \sg = \bigoplus_{k=1}^n \left(
\begin{array}{cccc} 0 & 1 \\  -1 & 0  \end{array} \right) \non \ee 

The displacement vector are irrelevant in the study of entanglement and are taken to be zero in our examples. The covariance matrix thus reduces to 
\be \gm_{j,k}=2\,\Re \Tr[\rho R_j R_k] \ee

Any real symmetric positive-definite matrix A can be brought to its Williamson normal form \cite{Will:1936} via symplectic transformations, i.e. transformations that preserve the canonical commutation relations, $A_{WF} = S A S ^T = \diag(a_1, a_1, .... a_n, a_n)$ where the $a_i$'s are the symplectic eigenvalues of A. One can calculate them as the positive eigenvalues of $\im \sg A$ or more simply as the positive square root of the eigenvalues of $-\sg A \sg A$.

A particularly suitable measure of the entanglement of mixed Gaussian states is the logarithmic negativity \cite{Werner:2002, Anders:2003, Eisert:2003, Eisert:2004}. It vanishes for separable states, does not increase under LOCC (local operations and classical communication), and stays invariant under local unitary transformations. It is defined as  
\be
\mathcal{E}_{\mathcal{N}}(\rho) = - \sum _{i=1} ^{2n} \log_2 \,
(\min\,(1\, ,\mid \lb_i \mid)), \ee 
where the $\lambda_i$ are the symplectic eigenvalues of the partially transposed covariance $\gamma^{(T_1)}$, which is obtained from $\gm$ by reversing the
time in all variables of one of the subsystems. Choosing to transpose with respect to particle 1, we replace $x_1 \rightarrow
x_1$ and $p_1 \rightarrow -p_1$. The $\lb_i$'s are thus the square roots of the eigenvalues of $- \sg \gm^{T_1} \sg \gm^{T_1}$.

\section{Free evolution of an entangled initial state\label{result}}

We first consider the case of a free particle Hamitonian 
\begin{equation} \label{hamilt1}
H = \frac{p_1 ^2}{2m} + \frac{p_2 ^2}{2m}
\end{equation}
with $p \vert x \rangle = - \im \hb \frac{\partial}{\partial x} \vert x \rangle$. This will allow us to examine the dynamics of the entanglement when an entangled bipartite Gaussian state is left to evolve, each particle coupled to its own heat bath. (\ref{mastereqn}) can be solved to obtain
\begin{eqnarray} \label{sol}
\lefteqn{{\tilde P} (\tbf{q}, \tbf{z}, t) = {\tilde P} (\tbf{q}, \tbf{z}_0(t), 0) \exp \left[ - \tau_2 \,q_2^2\, t - \tau_1 \,q_1
^2\, t \right]} \non \\ && \times \exp \biggl[  -\lb_1(t) \, \left( z_1 + \frac{q_1}{2\gm_1}\right) ^2 -\lb_2(t) \, \left( z_2
+ \frac{q_2}{2\gm_2}\right) ^2 \non 
\\ && \hspace{0.5 in} + \alp_1(t) \, \left( z_1 + \frac{q_1}{2\gm_1}\right) + \alp_2(t) \,
\left( z_1 + \frac{q_2}{2\gm_2}\right) \biggr] \non \\
\end{eqnarray}
with
\begin{equation}
\begin{array}{c}
\lb_i(t) =  2 m k T_i (1 - e^{-2\gm_i t/m}) \non
\\ \alp_i(t) = \frac{4 m k T_i}{\gm_i} (1 - e^{-\gm_i t/m})
\\ \tau_i = \frac{k T_i}{\gm_i}
\\ z_{0,i}(t) = z_i\, e^{-\gm_i t/m} - \frac{q_i}{2\gm_i} \,
\left(1 - e^{-\gm_i t/m}\right)
\end{array}
\end{equation}
The full derivation for this solution can be found in Appendix~\ref{solmasteq}.
Let us consider a bipartite initial state with the Gaussian wavefunction, suggested by Ford and O'Connell \cite{FO'C:2008, FO'C:2010} 
\be \label{initstate} \Psi(x_1, x_2) = \Og ^{1/2} e^{-\frac{(x_1 - x_2 ) ^2}{4 s ^2}} e^{-\frac{(x_1 + x_2)
^2}{16d ^2}} \ee
The corresponding density matrix is
\begin{equation} \label{rho0} \rho\big|_{t=0}= \Og e^{-\ep_{+} ({x_1}^2+{x_2}^2+{x'_1}^2+{x'_2}^2) + 2 \ep_{-} (x_1 x_2+x'_1 x'_2)} 
\end{equation} 
where $\Og=\frac{1}{2 \pi s d}$, $\ep_{\pm}=\frac{1}{4s ^2} \pm \frac{1}{16d ^2}$.

\sbp We next compute the time-evolved state by inserting the Fourier transform of (\ref{rho0}) into (\ref{sol}). The Fourier transform of (\ref{rho0}) is
\begin{eqnarray}
\lefteqn{{\tilde P} (\tbf{q},\tbf{z}_0;0) } \non
\\ &=& \exp\left[- 2\ep_+ \hb ^2 {z_0}_1^2
- 2\ep_{+} \hb ^2{z_0}_2 ^2 + 4 \ep_{-}\hb ^2 {z_0}_1 {z_0}_2\right] \non
\\ && \times \exp \left[-\frac{\ep_+ ({q_1}^2 + q_2^2)}{8 (\ep_{+}^2 - \ep_- ^2)}
- \frac{\ep_- q_1 q_2}{4 (\ep_{+}^2 - \ep_- ^2)}\right]
\end{eqnarray}
which, when substituted into (\ref{sol}) yields
\begin{eqnarray}  \label{tildePt}
{\tilde P}(\tbf{q},\tbf{z},t )&=& e^{- A_1 {q_1}^2 - A_2 {q_2}^2 - B_1 {z_1}^2 - B_2 {z_2}^2 - D z_1 z_2 - E q_1 q_2}\non
\\ && \times\,e^{- C_{11} z_1 q_1 - C_{22} z_2 q_2 - C_{12} z_1 q_2 - C_{21} z_2 q_1}, \end{eqnarray} 
where the coefficients are given by
\begin{eqnarray} \label{coeff}
 A_j &=& \frac{d ^2}{2} + \frac{s ^2}{8} + \tau _j - \frac{\alp _j}{2\gm _j}
 + \frac{\lb _j}{4{\gm _j}^2} + \frac{\hb ^2\ep_{+}}{2{\gm _j}^2}\,
 (1-e^{-\frac{\gm _j t}{m}})^2 \non
\\ B_j &=& 2 \hb ^2\ep_{+} \,e^{-\frac{2\gm _j t}{m}}+\lb _j  \non
\\ C_{jk} &=& \frac{2 \hb ^2\ep_{\pm}}{\gm _j}\, e^{-\frac{\gm _j t}{m}}\,
(1 - e^{-\frac{\gm _k t}{m}}) - \left(\frac{\lb _j}{\gm _j}+\alp
_j \right) \dt_{jk} \non \\ &&
\text{ with} \begin{cases} \ep _+ \, &\text{if $j = k$} \non \\
\ep_- \, &\text{if $j \neq k$} \end{cases}
\\  D &=& 2 \hb ^2\ep_{-}\,e^{-\frac{\gm _1 t}{m}}\,e^{-\frac{\gm _2 t}{m}} \non
\\  E &=& 2 d ^2 - \frac{s ^2}{2} -
\frac{2 \hb ^2\ep_{-}}{\gm _1 \gm _2}(1 - e^{-\frac{\gm _1 t}{m}})
(1 - e^{-\frac{\gm _2 t}{m}})\non \\
\end{eqnarray}

\sbp The entries of the covariance matrix can be calculated directly from (\ref{tildePt}) taking into account the change of variables (\ref{changevariable}):
\begin{eqnarray}
2\, \Rl \langle X_i X_j \rangle &=& - 2 \left( \frac{\partial}{\partial q_i} \frac{\partial}{\partial q_j} {\tilde P}(\tbf{q}, \tbf{z}=0, t)\right) \vert_{\tbf{q}=0} \non
\\ 2\, \Rl \langle X_i P_j \rangle &=& \frac{\partial}{\partial q_i} \frac{\partial}{\partial z_j}\, {\tilde P}(\tbf{q}, \tbf{z}, t)
\vert_{\tbf{q}=0, \tbf{z}=0}  \non
\\ 2\, \Rl \langle P_i X_j \rangle &=&  \frac{\partial}{\partial z_i} \frac{\partial}{\partial q_j}\, {\tilde P}(\tbf{q}, \tbf{z}, t)
\vert_{\tbf{q}=0, \tbf{z}=0} \non
\\ 2\, \Rl \langle P_i P_j \rangle &=& - \frac{1}{2} \left( \frac{\partial}{\partial z_j}\, \frac{\partial}{\partial z_j} \, {\tilde P}(\tbf{q}=0, \tbf{z}, t)\right) \vert_{\tbf{z}=0} \non
\end{eqnarray}
The covariance matrix is then 
\be \gm_t = \left(
\begin{array}{cccc}  4 A_1 & - C_{11} & E & - C_{21} \\ - C_{11} &
B_1 & - C_{12} & D \\ E & -C_{12} & 4 A_2 & - C_{22} \\ - C_{21} &
D & - C_{22} & B_2
\end{array} \right) \ee

We now perform the partial transposition with respect to particle 1: 
\be \gm_t ^{T_1} = \left(\begin{array}{cccc} 4 A_1 & C_{11} & E & - C_{21} 
\\ C_{11} & B_1 & C_{12} & -D \\ E & C_{12} & 4 A_2 & - C_{22} 
\\ - C_{21} & -D & - C_{22} & B_2 \end{array} \right)
\ee 
which is real and symmetric.

The symplectic eigenvalues are given by
\be
\lb_{\pm} ^T = \frac{\vep_{11} + \,\vep_{33}}{2} \pm \half \, \sqrt{ (\vep_{11} - \, \vep_{33}) ^2 + 4 \vep_{13} \vep_{24} - 4 \vep_{14} \vep_{23}}
\ee
where $\vep_{12} = \vep_{21} = \vep_{34} = \vep_{43} = 0$ and
\begin{eqnarray}
 \vep_{11} = \vep_{22} =& 4 A_1 B_1 - D E + C_{12} C_{21} -C_{11} ^2 \non
\\ \vep_{33} = \vep_{44} =& 4 A_2 B_2 - C_{22} ^2 - D E + C_{12} C_{21} \non
\\ \vep_{13} = \vep_{42} =& E B_1 - 4 A_2 D - C_{11} C_{12} + C_{12} C_{22} \non
\\ \vep_{14} = - \vep_{32}=& - C_{12} B_2 - C_{21} B_1 + C_{11} D + C_{22} D \non
\\ \vep_{23} = - \vep_{41} =& - E C_{11} + 4 A_1 C_{12} + 4 A_2 C_{21} - E C_{22} \non
\\ \vep_{24} = \vep_{31} =& E B_2 - C_{22} C_{21} + C_{11} C_{21} - 4 A_1 D  \non \\
\end{eqnarray}
The logarithmic negativity then becomes 
\be
\mathcal{E_{\mathcal{N}}}(\rho) = - 2 \left( \log_2\left(\min(1,\vert  \lb_{+}^T  \vert)\right) + \log_2\left(\min(1,\vert \lb_{-}^T  \vert)\right) \right) \ee

\begin{figure}
\begin{center}
    \includegraphics[scale=0.45]{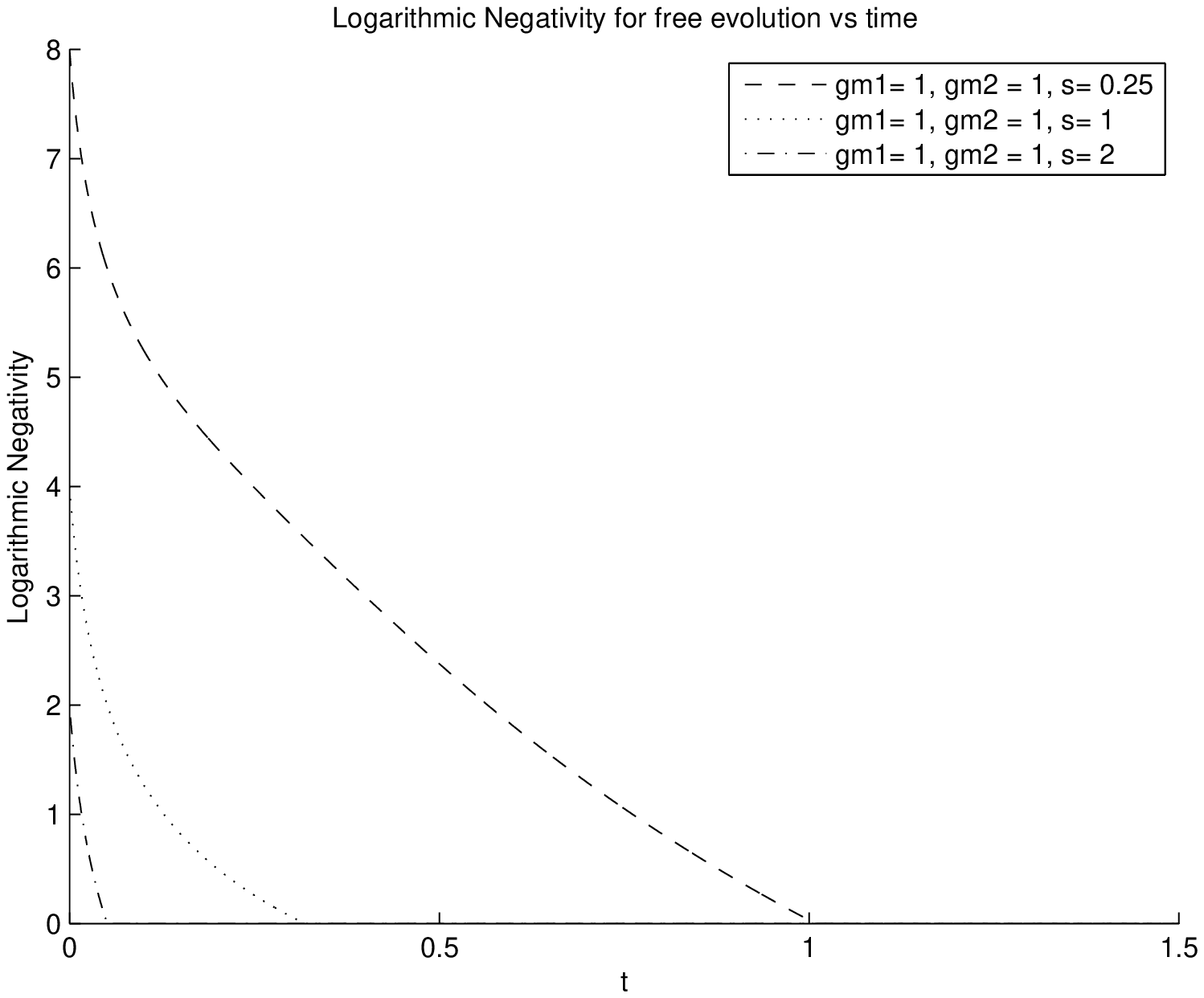}
    \caption{Logarithmic negativity vs t for three values of $s$ }
    \medskip{The values of $s$ are : dashed $s = 0.25$, dotted $s = 1$, dash-dotted $s = 2$}
\label{diff_sg1}
\end{center}
\end{figure}

Figure~\ref{diff_sg1} shows the logarithmic negativity as a function of time for three values of $s$. We can observe that there is complete disentanglement between the particles from a sharp cut-off time onwards, which obviously depends on $s$, and hence on the initial degree of entanglement. The sharp cut-off time characterizes entanglement sudden death (ESD). 

%include the bipharmpot stuff
\section{Evolution with a harmonic potential interaction\label{bipharmpot}}

If we introduce a harmonic potential interaction into the Hamiltonian, (\ref{hamilt1}) generalises to 
\be H_s = \frac{p_1 ^2}{2m} + \frac{p_2 ^2}{2m} + \frac{m \og_0 ^2}{2} (x_1 - x_2) ^2 \ee 
We can include this into (\ref{mastereqn}) and solve the resulting differential equation following the method described in appendix~\ref{solmasteq}. In this case \be M = \left( \begin{array}{cccc} 2\gm_1 & 0 & 1 & 0 
\\ 0 & 2\gm_2 & 0 & 1
 \\ - 4 m^2 \og_0 ^2 & 4 m^2 \og_0 ^2 & 0 & 0 
\\ 4 m^2 \og_0 ^2 & - 4 m^2 \og_0 ^2 & 0 & 0 \end{array} \right) \ee %
In general, the eigenvalue equation is quartic and the solution is complicated. We therefore assume, for simplicity, that $\gm_1 = \gm_2 = \gm$ and $T_1 = T_2 = T$. In that case the eigenvalues are
\be \label{Meigs} \bds{\lb} = \left(0, 2 \gm , \gm + \sqrt{\gm ^2
- 8 m^2 \og_0 ^2} , \gm - \sqrt{\gm ^2 - 8 m^2 \og_0 ^2} \right)^T
\ee 
and after some unpleasant algebra, we can write 
\bea
 {\tilde P} &=& \exp \bigl[- A q_1 ^2 - A q_2 ^2 - E q_1 q_2 - B z_1 ^2 - B z_2 ^2 - D z_1 z_2 \non
\\ && \hspace{0.1 in} - C_1 z_1 q_1 - C_1 z_2 q_2 - C_2 z_1 q_2 - C_2 z_2 q_1 \bigr] \non \\
\eea 
which is of the same form as (\ref{tildePt}), with $A_1 = A_2 = A$, $B_1 = B_2 = B$, $C_{11} = C_{22} = C_1$ and $C_{12} =
C_{21} = C_2$ except that the explicit expressions for $A$, $B$, etc. are more complicated, and will be omitted here. The following
figures show the logarithmic negativity for various values of $\og_0$ and of $\gm$.

\begin{figure}[h]
 \begin{center}
    \includegraphics[scale=0.45]{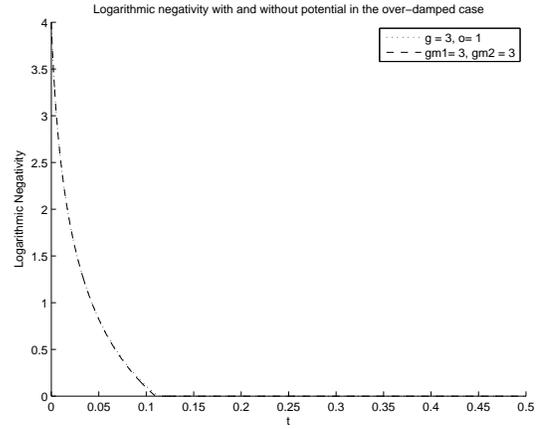}
    \caption{Logarithmic Negativities with and without potential for $\og_0 = 1$ and $\gm = \gm_1 = \gm_2 = 3$}
    \medskip{The dotted line represents $\mathcal{E}$ plotted with the potential.}
    \label{BLNover}
\end{center}
\end{figure}

\begin{figure}[h]
 \begin{center}
    \includegraphics[scale=0.45]{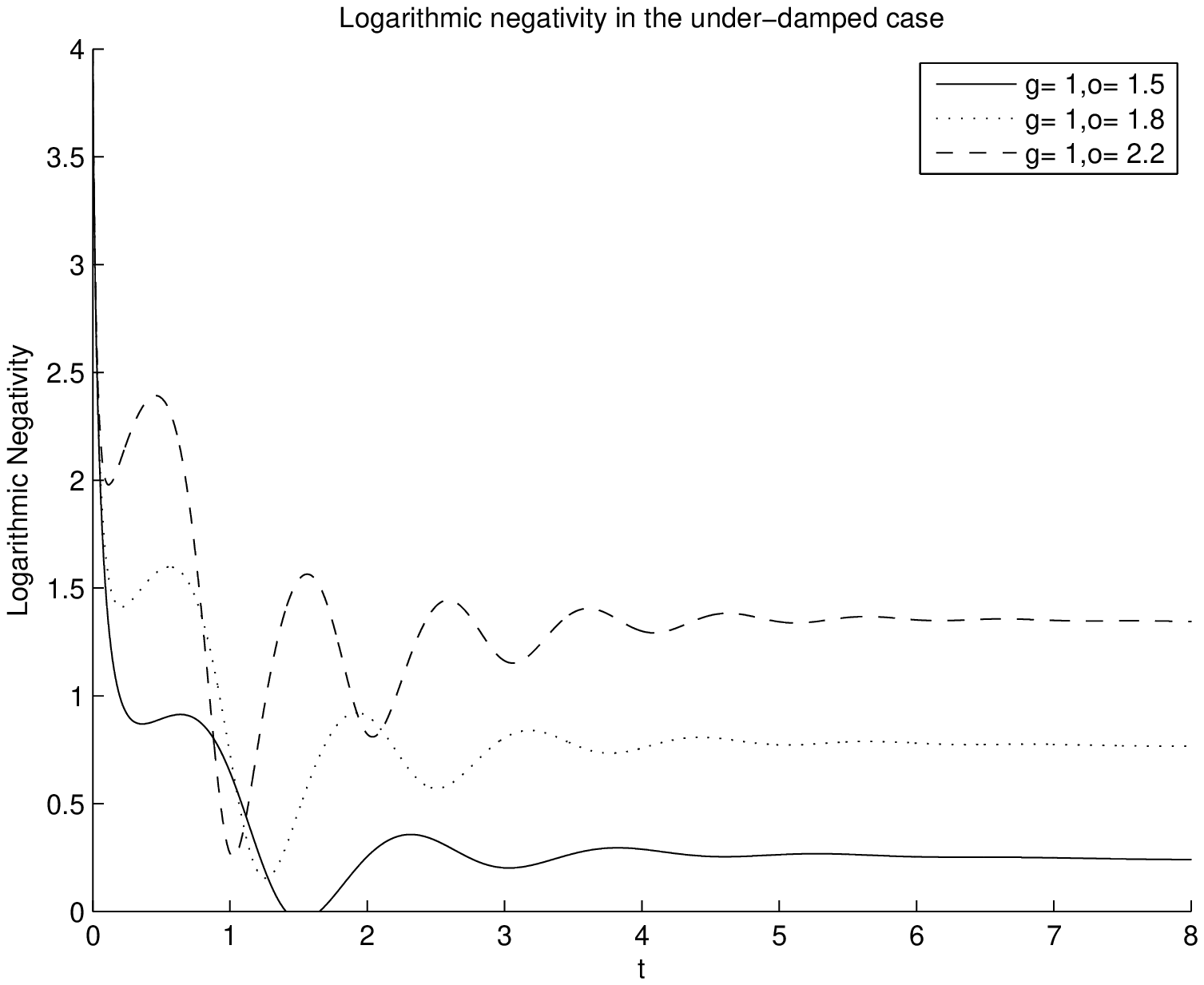}
    \caption{Logarithmic Negativities with and without potential for $\og_0 = 1$ and $\gm = \gm_1 = \gm_2 = 0.2$}
    \medskip{The dotted line represents $\mathcal{E}$ plotted with the potential.}
    \label{BLNunder}
\end{center}
\end{figure}

\begin{figure}[h]
 \begin{center}
    \includegraphics[scale=0.45]{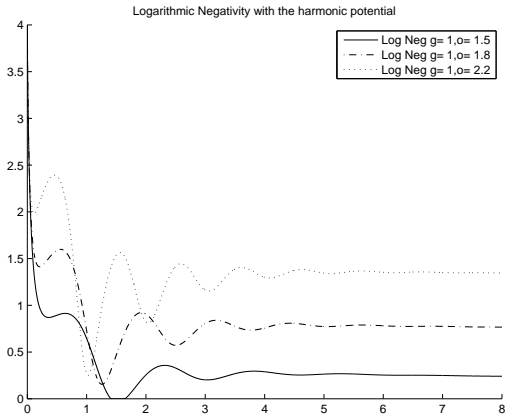}
    \caption{$\mathcal{E}$ in the highly under-damped case}
    \medskip{The plots are obtained with $\gm = 1$ and : full $\og_0 = 1.5$, dashed $\og_0 = 1.8$ and dotted $\og_0 = 2.2$}
    \label{VeryUD}
\end{center}
\end{figure}

\section{Concluding observations \label{conclusion}}

\sbp Figure~\ref{BLNover} and Figure~\ref{BLNunder} illustrate that, in the presence of a harmonic interaction between the particles, there is a marked difference in behaviour between two damping regimes. In the over-damped case ($\gm > 2 \sqrt{2} \og_0$), Figure~\ref{BLNover}, it can easily be seen that the curves coincide. This suggests that if the coupling is much stronger than the harmonic potential, the decay of the entanglement is unaffected by the potential. As a matter of fact, closer investigation of the vanishing point shows that the entanglement disappears faster as damping increasing. 
On the other hand, if the damping is small ($\gm < 2 \sqrt{2} \og_0$), the entanglement can reappear several times, as illustrated by Figure~\ref{BLNunder}. Figure~\ref{VeryUD} shows the behaviour of the entanglement as the damping decreases still. It can easily be seen that as the harmonic potential becomes stronger, the entanglement does not disappear. Instead it decreases sharply before being "restored". It then tends towards a non-zero constant value for large times. This suggests that allowing the particles to interact harmonically effectively saves the entanglement.
\medskip

\begin{acknowledgments}

This research was supported by a Dublin Institute for Advanced Studies scholarship. We would like to thank Prof. Robert O'Connell and Prof. Daniel Heffernan for their invaluable advice.
\end{acknowledgments}

\appendix
\section*{The Master equation \label{derivmasteq}}
In this section, we will describe the derivation of the master equation in more details. A simple derivation of the Quantum Langevin Equation, starting from a heat bath modelled by independent oscillators coupled harmonically to a system of one particle, was given in \cite{FLO'C:QLE1988}. The Hamiltonian has the form
\begin{equation}
H = \frac{p^2}{2m} + V(x) + \frac{1}{2} \sum_j \left\{
\frac{p_j^2}{m_j} + m_j \og_j ^2 (q_j - x)^2 \right\}
\end{equation}
Solving the Heisenberg equations of motion yields
\begin{eqnarray}
q_j (t) &=& q_j^h(t) + x(t) - \int_{-\infty} ^t \cos\left[\og_j
(t-t')\right] {\dot x} (t') \, dt' \non
\\ q_j^h (t) &=& q_j \cos(\og_j t) + \frac{p_j}{\og_j m_j} \sin (\og_j
t).
\end{eqnarray}
Introducing the quantities
\begin{eqnarray}
\mu (t) &=& \sum_j m_j \og_j ^2 \cos(\og_j t) \Theta(t) \non
\\ \xi (t) &=& \sum_j m_j \og_j ^2 q_j ^h (t)
\end{eqnarray}
($\Theta(t)$ is the Heaviside function) we obtain the Quantum Langevin Equation
\begin{eqnarray}
m {\ddot x}_k + \int_{-\infty} ^t \mu_k (t-t') {\dot x}_k(t')\, dt' + V'(x_k) &=& \xi_k(t)
\end{eqnarray}
where the dot denotes the derivative with respect to time and the prime that with respect to $x$. $\mu(t)$ and $\xi(t)$ describe the influence of the bath on the system and are known as the \textit{memory function} and the operator-valued \textit{random force} respectively. We also introduce the \textit{spectral distribution}
\begin{eqnarray}
 G(\og) &=&  \Re \left[ {\tilde \mu} (\og + \im 0 ^+)\right] \non 
\\ &=& \frac{\pi}{2} \sum_j m_j \og_j^2 \left[ \dt(\og - \og_j) + \dt( \og + \og_j)\right],
\end{eqnarray}
in terms of which the autocorrelation of $\xi(t)$ is given by
\begin{eqnarray} \label{autocor}
 \lefteqn{\frac{1}{2} \langle \left[ \xi(t), \xi(t') \right]_+ \rangle =}
 \non \\ && \frac{1}{\pi} \int_0^{\infty} G(\og) \hb \og \coth
 \left( \frac{\hb \og}{2 k T} \right) \cos \left[\og (t-t') \right] \, d\og
\end{eqnarray}
where $[ \, , \, ]_+$ denotes the anticommutator. For a general observable $Y$ of the small system (particle), one can write
\begin{eqnarray}
{\dot Y} &=& \frac{\im}{\hb} \left[ H, Y \right] \non 
\\ &=& \frac{\im}{\hb} \left[ H_s, Y \right] - \frac{\im}{2\hb}
\left[ \left[ x ,Y \right], \xi(t) \right]_+ \non 
\\ && + \frac{\im}{2\hb} \left[ \left[ x ,Y \right], \int_{-\infty}^t
dt'\,\mu(t-t') {\dot x}(t') \right]_+ \non \\ &&
\end{eqnarray}
In the case of an Ohmic heat bath, we can replace
\begin{equation}
 \int_{-\infty} ^t \mu(t') {\dot x}(t') \, dt' \rightarrow \gm {\dot x}(t)
 \qquad \text{and} \qquad G(\og) \rightarrow \gm
\end{equation}
so that the Quantum Langevin Equation reads
\begin{equation} \label{qle1} {\dot Y} = \frac{\im}{\hb}
\left[ H_s, Y \right] - \frac{\im}{2\hb} \left[ \left[ x ,Y
\right], \xi (t) \right]_+ + \frac{\im \gm}{2\hb} \left[ \left[
x, Y \right], {\dot x} (t) \right]_+. \end{equation} 
For a system of two particles, each connected to their individual heat bath, we have analogously
\begin{eqnarray} \label{qle}
{\dot Y} &=& \frac{\im}{\hb} \left[ H_s, Y \right] \non 
\\ && - \frac{\im}{2\hb} \left[ \left[ x_1 ,Y \right], \xi_1 (t)
\right]_+ - \frac{\im}{2\hb} \left[ \left[ x_2 ,Y \right],
\xi_2 (t) \right]_+ \non \\ && + \frac{\im}{2\hb} \left[ \left[
x_1 , Y \right] , \gm_1 {\dot x_1} (t) \right]_+ +
\frac{\im}{2\hb} \left[ \left[ x_2 , Y \right] ,\gm_2 {\dot
x_2} (t) \right]_+ \non \\
\end{eqnarray}
If we define
\begin{equation} \label{trs}
 \Tr_s \left\{ Y(t) \rho \right\} = \Tr_s \left\{ Y \rho(t) \right\}
\end{equation}
where $\Tr_s$ is the trace over the system. Let us introduce $\rho(t) = \rho_s(t) \otimes \rho_B$ where $\rho_s$ is the density matrix of the system and $\rho_B$ that of the bath. It follows easily from (\ref{qle}) that $\rho(t)$ satisfies the adjoint equation
\begin{eqnarray} \label{adj}
{\dot \rho}(t) &=& - \frac{\im}{\hb} \left[ H_s , \rho(t) \right]
- \frac{\im}{2\hb} \left[ \left[ \xi_1 (t) , \rho(t) \right]_+ ,
x_1 \right] \non \\ &&  - \frac{\im}{2\hb} \left[ \left[ \xi_2
(t) , \rho(t) \right]_+ , x_2 \right] + \frac{\im\gm_1}{2\hb}
\left[ \left[ {\dot x}_1 , \rho(t) \right]_+ , x_1 \right]  \non \\
&& + \frac{\im\gm_2}{2\hb} \left[ \left[ {\dot x}_2 , \rho(t)
\right]_+ , x_2 \right]
\end{eqnarray}
To derive the master equation (that is, an effective equation for $\rho_s(t)$) from the adjoint equation, we assume that the noise is small and temporarily introduce a small parameter $\ep$, replacing $\xi(t)$ by $\ep \xi(t)$. We can write $\nu(t)$ to second order in $\ep$ as
\begin{equation}
\nu(t) = \nu_0(t) + \ep \nu_1(t) + \ep ^2 \nu_2 (t) \non
\end{equation}
We also assume the baths and the system are decoupled at $t=-\infty$, so that $\rho_0(t) = \rho_0(t) \rho_B$. Inserting this expansion into (\ref{adj}) yields equations for $\rho_0$, $\rho_1$ and $\rho_2$ which can be solved successively. The equation for $\rho_0$ reads 
\begin{eqnarray} {\dot \rho}_0(t) &=&
-\frac{\im}{\hb} [H_s,\rho_0(t)] \non 
\\ && + \frac{\im \gm _1}{2 \hb}
\left[ \left[ {\dot x}_1, \rho_0(t)\right]_+, x_1 \right] +
\frac{\im \gm _2}{2 \hb} \left[ \left[ {\dot x}_2,
\rho_0(t)\right]_+, x_2 \right]. \non \\
\end{eqnarray}
The equation for $\rho_1$ can be written as
\begin{eqnarray} {\dot \rho}_1(t) &=&
-\frac{\im}{\hb} [H_s,\rho_1(t)] \non 
\\ && + \frac{\im \gm _1}{2 \hb}
\left[ \left[ {\dot x}_1, \rho_1(t)\right]_+, x_1 \right] - \frac{\im}{2 \hb} [\xi_1(t),\rho_B]_+ [\rho_0(t),x_1] \non 
\\ && + \frac{\im \gm _2}{2 \hb} \left[ \left[ {\dot x}_2, \rho_1(t)\right]_+, x_2 \right] - \frac{\im}{2 \hb}
[\xi_2(t),\rho_B]_+ [\rho_0(t),x_2]. \non \\
\end{eqnarray} 
The solution for the first-order term can be written as
\begin{eqnarray}
\rho_1 (t) &=& - \frac{\im}{2\hb} \int_{-\infty} ^t e^{A_s
(t-t')} \Big\{ \left[ \rho_0 (t'), x_1 \right] \otimes \left[
\xi_1(t'), \rho_B \right]_+ \non \\ && \hspace{1 in} + \left[
\rho_0 (t'), x_2 \right] \otimes \left[ \xi_2(t'), \rho_B
\right]_+ \Big\} \, dt' \non
\end{eqnarray}
where $A_s$ is a super operator which, applied to $\rho_k(t)$ yields
\begin{eqnarray}
\lefteqn{A_s \rho_k(t) = -\frac{\im}{\hb} \left[H_s ,
\rho_k(t)\right]} \non \\ && + \frac{\im \gm_1}{2\hb} \left[
\left[ {\dot x}_1 , \rho_k(t) \right]_+ , x_1 \right] +
\frac{\im \gm_2}{2\hb}
\left[ \left[ {\dot x}_2 , \rho_k(t) \right]_+ , x_2 \right]. \non \\
\end{eqnarray}
Finally, we insert this solution into the equation for $\rho_2$:
\begin{eqnarray} {\dot \rho}_2(t) &=& -\frac{\im}{\hb} [H_s,\rho_2(t)] \non 
\\ && + \frac{\im \gm_1}{2 \hb}
\left[ \left[ {\dot x}_1, \rho_2(t)\right]_+, x_1 \right] +
\frac{\im \gm_2}{2 \hb} \left[ \left[ {\dot x}_2,
\rho_2(t)\right]_+, x_2 \right] \non 
\\ && - \frac{\im}{2 \hb}
\left[ [\xi_1(t),\rho_1(t)]_+, x_1 \right] - \frac{\im}{2 \hb}
\left[ [\xi_2(t),\rho_1(t)]_+, x_2 \right] \non \\
\end{eqnarray}
Taking the trace over the bath variables and using the autocorrelation (\ref{autocor}) in the Ohmic limit, we obtain the non-rotating-wave master equation (upon removal of $\ep$)
%\begin{widetext}
\begin{eqnarray} 
{\dot \rho_s}(t) &=& -\frac{\im}{\hb} \left[H_s , \rho_s (t)\right] \non
\\ && + \frac{\im \gm_1}{2\hb} \left[ \left[ {\dot x}_1 , \rho_s (t) \right]_+ , x_1 \right] 
+ \frac{\im \gm_2}{2\hb} \left[ \left[ {\dot x}_2 , \rho_s (t) \right]_+ , x_2 \right] \non 
\\ && - \frac{k T_1 \gm_1}{\hb ^2} \left[ \left[ \rho_s(t) , x_1 \right] , x_1 \right]
 - \frac{k T_2 \gm_2}{\hb ^2} \left[ \left[ \rho_s(t) , x_2 \right] , x_2 \right]. \non \\
\end{eqnarray}
%\end{widetext}

\section*{Solution to the master equation \label{solmasteq}}
The derivation of the solution to the master equation will here be given in a general way. The method will be described for a free particle system Hamiltonian 
\begin{equation}
H = \frac{p_1 ^2}{2m} + \frac{p_2 ^2}{2m}
\end{equation}
with $p \vert x \rangle = - \im \hb \frac{\partial}{\partial x} \vert x \rangle$, but is easily generalised to other types of Hamiltonians. Note that we assume that the particles have the same mass. In position-space, (\ref{mastereqn}) becomes
%\begin{widetext}
\begin{eqnarray} \lefteqn{\frac{\partial}{\partial t} \langle x|\,\rho\,|y\rangle =} \non
\\ && \frac{\im \hb}{2 m} \left( \frac{\partial ^2}{\partial x_1 ^2} - \frac{\partial ^2}{\partial y_1 ^2} + \frac{\partial ^2}{\partial x_2 ^2} -\frac{\partial ^2}{\partial y_2 ^2} \right) \rho \non 
\\ && - \frac{\gm_1}{2 m} \left( x_1 - y_1 \right) \left( \frac{\partial }{\partial x_1} - \frac{\partial}{\partial y_1} \right) \rho \non 
\\ && - \frac{\gm_2}{2 m} \left( x_2 - y_2 \right) \left( \frac{\partial }{\partial x_2} - \frac{\partial}{\partial y_2} \right) \rho 
\\ && - \frac{\gm_1 k T_1}{\hb ^2} (x_1 - y_1)^2 \rho - \frac{\gm_2 k T_2}{\hb^2} (x_2 - y_2)^2 \rho \non
\end{eqnarray} 
%\end{widetext}
Using the change of variables 
\be \label{changevariable} x = u + \hb z \quad \text{and} \quad x' = u - \hb z \ee 
and replacing $\rho(\tbf{x}, \tbf{x'},0) \rightarrow P(\tbf{u}, \tbf{z},0)$, we apply a Fourier transformation with respect to $\tbf{u}$:
\begin{equation} \label{ftsol} {\tilde P}(\tbf{q}, \tbf{z}, t) =
\frac{1}{4\pi ^2} \int P(\tbf{u}, \tbf{z}, t) e^{-\im q_1 u_1 -
\im q_2 u_2} du_1\, du_2
\end{equation}
obtaining an equation for ${\tilde P}(\tbf{q}, \tbf{z}, t)$:
\begin{eqnarray} \lefteqn{\frac{\partial}{\partial t} {\tilde P}(\tbf{q},
\tbf{z}, t) =} \non 
\\ && -\left[ \left( \frac{\gm_1}{m} z_1 + \frac{q_1}{2
m} \right) \frac{\partial}{\partial z_1} + 4 \gm _1 k T_1 z_1^2
\right] {\tilde P}(\tbf{q}, \tbf{z}, t)  \\ && -\left[ \left(
\frac{\gm_2}{m} z_2 + \frac{q_2}{2 m} \right)
\frac{\partial}{\partial z_2} + 4 \gm _2 k T_2 z_2^2 \right] {\tilde
P}(\tbf{q}, \tbf{z}, t). \end{eqnarray} 
This equation can in principle again be solved using the method of characteristics. The characteristic equation is 
\be \frac{\partial \tbf{v}}{\partial t} = \frac{M}{2m} \tbf{v} \ee 
with $\tbf{v} = (z_1, z_2, q_1, q_2)^T$ and 
\be M = \left( \begin{array}{cccc} 2\gm_1 & 0 & 1 & 0 
\\ 0 & 2 \gm_2 & 0 & 1
 \\ 0 & 0 & 0 & 0 
\\ 0 & 0 & 0 & 0 \end{array} \right) \ee %
On a characteristic,
\bea \lefteqn{\frac{d}{dt} {\tilde P}(\tbf{q}, \tbf{z}(t), t)} \non 
\\ &=& -[4 \gm_1 k T_1 z_1 ^2(t) + 4 \gm_2 k T_2 z_2 ^2(t)]\, {\tilde P}(\tbf{q}, \tbf{z}(t), t) \non \eea 
The eigenvalues and eigenvectors of $M$ can be computed to be 
\begin{equation} \bds{\lb}^T = (2 \gm_1, 2\gm_2 , 0, 0) = (\lb_1, \lb_2, \lb_3, \lb_4)
\end{equation}
and \begin{equation} Q = \left( \begin{array}{cccc} 1 & 0 & -\frac{1}{2 \gm_1}  & 0 
\\ 0 & 1 & 0 & -\frac{1}{2 \gm_2}
\\ 0 & 0 & 1 & 0 \\ 0 & 0 & 0 & 1\end{array} \right)
\end{equation}
Since $Q^{-1} M Q = D$ where $D$ is the diagonal matrix, we need $Q^{-1}$ as
\begin{equation}
Q^{-1} = \left( \begin{array}{cccc} 1 & 0 & 0 & \frac{1}{2 \gm_1}
\\ 0 & 1 & \frac{1}{2 \gm_2} & 0 
\\ 0 & 0 & 1 & 0 \\ )0 & 0 & 0 & 1 \end{array} \right)
\end{equation}
Then we can write $2 m \frac{\ptl \tbf{w}}{\ptl t} = D \tbf{w} $ with $\tbf{w} = Q^{-1} \tbf{v}$ which is easily solved so that $\tbf{v}(t) = Q e^{Dt / 2m} Q^{-1} \tbf{v}_0$ with
\begin{equation} e^{Dt/2m} = \left( \begin{array}{cccc} e^{\gm_1 t / m} & 0 & 0 & 0
\\ 0 & e^{\gm_2 t /m} & 0 & 0 \\ 0 & 0 & 0 & 0 \\ 0 & 0 & 0 & 0
\end{array} \right)
\end{equation}
Some more algebra yields the solution
\begin{eqnarray} 
\lefteqn{{\tilde P} (\tbf{q}, \tbf{z}, t) = {\tilde P} (\tbf{q}, \tbf{z}_0(t), 0) \exp \left[ - \tau_2 \,q_2^2\, t - \tau_1 \,q_1
^2\, t \right]} \non \\ && \times \exp \biggl[  -\lb_1(t) \, \left( z_1 + \frac{q_1}{2\gm_1}\right) ^2 -\lb_2(t) \, \left( z_2
+ \frac{q_2}{2\gm_2}\right) ^2 \non 
\\ && \hspace{0.5 in} + \alp_1(t) \, \left( z_1 + \frac{q_1}{2\gm_1}\right) + \alp_2(t) \,
\left( z_1 + \frac{q_2}{2\gm_2}\right) \biggr] \non \\
\end{eqnarray}
with
\begin{equation}
\begin{array}{c}
\lb_i(t) =  2 m k T_i (1 - e^{-2\gm_i t/m}) \non
\\ \alp_i(t) = \frac{4 m k T_i}{\gm_i} (1 - e^{-\gm_i t/m})
\\ \tau_i = \frac{k T_i}{\gm_i}
\\ z_{0,i}(t) = z_i\, e^{-\gm_i t/m} - \frac{q_i}{2\gm_i} \,
\left(1 - e^{-\gm_i t/m}\right)
\end{array}
\end{equation}
We thus have an expression giving the time dependency for an arbitrary initial state.

\bibliography{biblidraft1.bib}

\begin{thebibliography}{22}
\expandafter\ifx\csname natexlab\endcsname\relax\def\natexlab#1{#1}\fi
\expandafter\ifx\csname bibnamefont\endcsname\relax
  \def\bibnamefont#1{#1}\fi
\expandafter\ifx\csname bibfnamefont\endcsname\relax
  \def\bibfnamefont#1{#1}\fi
\expandafter\ifx\csname citenamefont\endcsname\relax
  \def\citenamefont#1{#1}\fi
\expandafter\ifx\csname url\endcsname\relax
  \def\url#1{\texttt{#1}}\fi
\expandafter\ifx\csname urlprefix\endcsname\relax\def\urlprefix{URL }\fi
\providecommand{\bibinfo}[2]{#2}
\providecommand{\eprint}[2][]{\url{#2}}

\bibitem[{\citenamefont{Podolsky and Rosen}(1935)}]{EPR:1935}
\bibinfo{author}{\bibfnamefont{A.~E.~B.} \bibnamefont{Podolsky}}
  \bibnamefont{and} \bibinfo{author}{\bibfnamefont{N.}~\bibnamefont{Rosen}},
  \bibinfo{journal}{Physical Review} \textbf{\bibinfo{volume}{47}},
  \bibinfo{pages}{777} (\bibinfo{year}{1935}).

\bibitem[{\citenamefont{Schr\"{o}dinger}(1935)}]{Schrod:1935}
\bibinfo{author}{\bibfnamefont{E.}~\bibnamefont{Schr\"{o}dinger}},
  \bibinfo{journal}{Proceedings of the Cambridge Philosophical Society}
  \textbf{\bibinfo{volume}{31}}, \bibinfo{pages}{555} (\bibinfo{year}{1935}).

\bibitem[{\citenamefont{Munro and Gardiner}(1996)}]{Gard:NRWME}
\bibinfo{author}{\bibfnamefont{W.}~\bibnamefont{Munro}} \bibnamefont{and}
  \bibinfo{author}{\bibfnamefont{C.}~\bibnamefont{Gardiner}},
  \bibinfo{journal}{Physical Review A} \textbf{\bibinfo{volume}{53}},
  \bibinfo{pages}{4} (\bibinfo{year}{1996}).

\bibitem[{\citenamefont{Gardiner and Zoller}(2000)}]{Gard:2000}
\bibinfo{author}{\bibfnamefont{C.}~\bibnamefont{Gardiner}} \bibnamefont{and}
  \bibinfo{author}{\bibfnamefont{P.}~\bibnamefont{Zoller}},
  \emph{\bibinfo{title}{Quantum Noise}} (\bibinfo{publisher}{Springer},
  \bibinfo{year}{2000}), \bibinfo{edition}{2nd} ed.

\bibitem[{\citenamefont{Lewis and O'Connell}(1988)}]{FLO'C:QLE1988}
\bibinfo{author}{\bibfnamefont{G.~F.~J.} \bibnamefont{Lewis}} \bibnamefont{and}
  \bibinfo{author}{\bibfnamefont{R.}~\bibnamefont{O'Connell}},
  \bibinfo{journal}{Physical Review A} \textbf{\bibinfo{volume}{37}},
  \bibinfo{pages}{11} (\bibinfo{year}{1988}).

\bibitem[{\citenamefont{Ghesqui\`{e}re}(2009)}]{Sakuya:2009}
\bibinfo{author}{\bibfnamefont{A.}~\bibnamefont{Ghesqui\`{e}re}}, Ph.D. thesis
  (\bibinfo{year}{2009}).

\bibitem[{\citenamefont{Yu and Eberly}(2003)}]{YE:2003}
\bibinfo{author}{\bibfnamefont{T.}~\bibnamefont{Yu}} \bibnamefont{and}
  \bibinfo{author}{\bibfnamefont{J.~H.} \bibnamefont{Eberly}},
  \bibinfo{journal}{Physical Review B} \textbf{\bibinfo{volume}{68}},
  \bibinfo{pages}{165322} (\bibinfo{year}{2003}).

\bibitem[{\citenamefont{Yu and Eberly}(2004)}]{YE:2004}
\bibinfo{author}{\bibfnamefont{T.}~\bibnamefont{Yu}} \bibnamefont{and}
  \bibinfo{author}{\bibfnamefont{J.~H.} \bibnamefont{Eberly}},
  \bibinfo{journal}{Physical Review Letters} \textbf{\bibinfo{volume}{93}},
  \bibinfo{pages}{14} (\bibinfo{year}{2004}).

\bibitem[{\citenamefont{Yu and Eberly}(2006)}]{YE:2006}
\bibinfo{author}{\bibfnamefont{T.}~\bibnamefont{Yu}} \bibnamefont{and}
  \bibinfo{author}{\bibfnamefont{J.~H.} \bibnamefont{Eberly}},
  \bibinfo{journal}{Physical Review Letters} \textbf{\bibinfo{volume}{97}},
  \bibinfo{pages}{140403} (\bibinfo{year}{2006}).

\bibitem[{\citenamefont{Pratt and Eberly}(2001)}]{PE:2001}
\bibinfo{author}{\bibfnamefont{J.~S.} \bibnamefont{Pratt}} \bibnamefont{and}
  \bibinfo{author}{\bibfnamefont{J.~H.} \bibnamefont{Eberly}},
  \bibinfo{journal}{Physical Review B} \textbf{\bibinfo{volume}{64}},
  \bibinfo{pages}{195314} (\bibinfo{year}{2001}).

\bibitem[{\citenamefont{Di\'{o}si}(2003)}]{Diosi:2003}
\bibinfo{author}{\bibfnamefont{L.}~\bibnamefont{Di\'{o}si}},
  \bibinfo{journal}{LANL e-print quant-ph/0301096}  (\bibinfo{year}{2003}).

\bibitem[{\citenamefont{Roszak and Machnikowski}(2006)}]{RM:2006}
\bibinfo{author}{\bibfnamefont{K.}~\bibnamefont{Roszak}} \bibnamefont{and}
  \bibinfo{author}{\bibfnamefont{P.}~\bibnamefont{Machnikowski}},
  \bibinfo{journal}{Physical Review A} \textbf{\bibinfo{volume}{73}},
  \bibinfo{pages}{022313} (\bibinfo{year}{2006}).

\bibitem[{\citenamefont{Lindblad}(1976)}]{Lindblad:1976}
\bibinfo{author}{\bibfnamefont{G.}~\bibnamefont{Lindblad}},
  \bibinfo{journal}{Communications in Mathematical Physics}
  \textbf{\bibinfo{volume}{48}}, \bibinfo{pages}{119} (\bibinfo{year}{1976}).

\bibitem[{\citenamefont{Ficek and Tan\'{a}s}(2006)}]{Ficek:2006}
\bibinfo{author}{\bibfnamefont{Z.}~\bibnamefont{Ficek}} \bibnamefont{and}
  \bibinfo{author}{\bibfnamefont{R.}~\bibnamefont{Tan\'{a}s}},
  \bibinfo{journal}{Physical Review A} \textbf{\bibinfo{volume}{74}},
  \bibinfo{pages}{024304} (\bibinfo{year}{2006}).

\bibitem[{\citenamefont{Ficek and Tan\'{a}s}(2008)}]{Ficek:2008}
\bibinfo{author}{\bibfnamefont{Z.}~\bibnamefont{Ficek}} \bibnamefont{and}
  \bibinfo{author}{\bibfnamefont{R.}~\bibnamefont{Tan\'{a}s}},
  \bibinfo{journal}{Physical Review A} \textbf{\bibinfo{volume}{77}},
  \bibinfo{pages}{054301} (\bibinfo{year}{2008}).

\bibitem[{\citenamefont{Anders}(2003)}]{Anders:2003}
\bibinfo{author}{\bibfnamefont{J.}~\bibnamefont{Anders}},
  \bibinfo{journal}{LANL e-print quant-ph/0610263}  (\bibinfo{year}{2003}).

\bibitem[{\citenamefont{Eisert and Plenio}(2003)}]{Eisert:2003}
\bibinfo{author}{\bibfnamefont{J.}~\bibnamefont{Eisert}} \bibnamefont{and}
  \bibinfo{author}{\bibfnamefont{M.}~\bibnamefont{Plenio}},
  \bibinfo{journal}{International Journal of Quantum Information}
  \textbf{\bibinfo{volume}{1}}, \bibinfo{pages}{479} (\bibinfo{year}{2003}).

\bibitem[{\citenamefont{Hartley and Eisert}(2004)}]{Eisert:2004}
\bibinfo{author}{\bibfnamefont{M.~P.~J.} \bibnamefont{Hartley}}
  \bibnamefont{and} \bibinfo{author}{\bibfnamefont{J.}~\bibnamefont{Eisert}},
  \bibinfo{journal}{New Journal of Physics} \textbf{\bibinfo{volume}{6}},
  \bibinfo{pages}{36} (\bibinfo{year}{2004}).

\bibitem[{\citenamefont{Williamson}(1936)}]{Will:1936}
\bibinfo{author}{\bibfnamefont{J.}~\bibnamefont{Williamson}},
  \bibinfo{journal}{American Journal of Mathematics}
  \textbf{\bibinfo{volume}{58}}, \bibinfo{pages}{141} (\bibinfo{year}{1936}).

\bibitem[{\citenamefont{Vidal and Werner}(2002)}]{Werner:2002}
\bibinfo{author}{\bibfnamefont{G.}~\bibnamefont{Vidal}} \bibnamefont{and}
  \bibinfo{author}{\bibfnamefont{R.}~\bibnamefont{Werner}},
  \bibinfo{journal}{Physical Review A} \textbf{\bibinfo{volume}{65}},
  \bibinfo{pages}{3} (\bibinfo{year}{2002}).

\bibitem[{\citenamefont{Ford and O'Connell}(2008)}]{FO'C:2008}
\bibinfo{author}{\bibfnamefont{G.~W.} \bibnamefont{Ford}} \bibnamefont{and}
  \bibinfo{author}{\bibfnamefont{R.~F.} \bibnamefont{O'Connell}},
  \bibinfo{journal}{Private Communications}  (\bibinfo{year}{2008}).

\bibitem[{\citenamefont{Gao and O'Connell}(2010)}]{FO'C:2010}
\bibinfo{author}{\bibfnamefont{G.~W. F.~Y.} \bibnamefont{Gao}}
  \bibnamefont{and} \bibinfo{author}{\bibfnamefont{R.~F.}
  \bibnamefont{O'Connell}}, \bibinfo{journal}{Optics Communications}
  \textbf{\bibinfo{volume}{283}}, \bibinfo{pages}{831} (\bibinfo{year}{2010}).

\end{thebibliography}

\end{document}